\documentclass{l4dc2027}
 
 \newtheorem{ass}{Assumption}[section]
 \newtheorem{prop}{Proposition}[section]
 \newtheorem{lem}{Lemma}[section]

\usepackage{enumitem}

\title[Control-Gibbs Distribution]{Control-Gibbs Distribution: \\ Sampling for Optimal Control and Diverse Actions}
\usepackage{times}

\coltauthor{\Name{Shengling Shi} \Email{Shengling.Shi@tudelft.nl}\\
 \addr Delft Center for Systems and Control, Delft University of Technology}

\begin{document}

\maketitle

\begin{abstract}%
While conventional optimal control selects an optimal input sequence, probabilistic control assigns probability mass over all possible inputs and exploits a distribution of control sequences. Control distributions help enable exploration, capture diverse solutions, and solve deterministic optimal control problems. This work reviews an important probabilistic model for control distributions, the control-Gibbs distribution, i.e., the Gibbs distribution with an optimal control objective as the energy function. We point out its fundamental role in various probabilistic control frameworks, e.g., maximum entropy control, control-as-inference framework, model predictive path integral control, and energy-based models for control. This provides a model-central perspective to unify these probabilistic control frameworks. We also review the associated extensions of dynamic programming for computing closed-loop control distributions. To sample the control-Gibbs distribution, we consider projected Langevin sampling and develop non-asymptotic guarantees. The guarantees show the dual benefits of sampling the control-Gibbs distribution: obtaining diverse control solutions and solving the deterministic optimal control problem. We provide a numerical example where diverse arm trajectories are sampled from a control-Gibbs distribution.
\end{abstract}


\section{Introduction}
Classical deterministic and stochastic optimal control typically seek an optimal input sequence or a deterministic feedback control law. Randomness often arises from uncertainties or noise and is regarded as an additional complication in the control problem. On the other hand, a probabilistic distribution of control input, control distribution in short, can be of interest, even if the system-to-controlled is deterministic. The randomness of control is intentionally introduced for problem solving. One motivation is to inject excitation, either for exploring better solutions as in reinforcement learning (RL) \citep{levine2018reinforcement} or for data informativity as in closed-loop identification \citep{forssell1999closed,athrey2024regret}. Another motivation is to model behavior variability in robotics \citep{chi2025diffusion} or neuroscience \citep{dhawale2017role}. For example, generative models are used as control distributions to imitate diverse human demonstrations \citep{urain2025survey}. 

To study control distributions theoretically, a common way is to start from a control problem formulation involving a probabilistic controller. \citet{karny1996towards,lefebvre2024probabilistic} study probabilistic control for distribution matching. That is, finding a control distribution that minimizes the KL divergence between the closed-loop trajectory distribution and a target distribution. This connects to entropy-regularized control problems \citep{bhole2025unifying,lambert2026entropy}, due to the cross-entropy decomposition of KL divergence. In the control-as-inference framework \citep{levine2018reinforcement}, a different derivation of a control distribution is given via probabilistic inference. There, an artificial optimality variable is introduced, and a control distribution is the posterior distribution of control given states and the optimality variable. Another related framework is the path integral control \citep{kappen2007introduction,kappen2016adaptive} and model predictive path integral control (MPPI) \citep{williams2018information}. Due to the imposed problem structure, e.g., additive noise on inputs \citep{williams2018information} or a special cost function \citep{kappen2007introduction,kappen2016adaptive}, a control distribution appears in the derivation and serves an intermediate role in deriving the final optimal control. 

All of these frameworks use a control distribution and share similarities, e.g., a KL formulation underlying distribution matching, control-as-inference, and path integral control. There have been efforts to connect or unify the above frameworks \citep{lefebvre2024probabilistic,bhole2025unifying}. However, due to different problem formulations and derivations, the discussions are hindered by technical details. In this work, we do not unify problem formulations, as one mathematical object can naturally be derived from different settings. Instead, we take a perspective of a modeling choice. We focus on the common model of a control distribution to draw connections between different frameworks. For simplicity, we consider the setting of a deterministic dynamical system.

 We review a special distribution of control: the control-Gibbs distribution with density $p(\mathbf{u}) \propto \text{exp}(-J(\mathbf{u})/\gamma)$, where $J(\mathbf{u})$ is a standard optimal control objective of an input sequence $\mathbf{u}$ and $\gamma>0$ is a parameter. A control sequence with a lower cost $J(\mathbf{u})$ has a higher probability. Note that for deterministic systems, $\text{exp}(-J(\mathbf{u})/\gamma)$ induces a trajectory distribution. In this work, we regard the control-Gibbs distribution as a canonical model for a control distribution.
 
 The control-Gibbs distribution is of independent interest for the following reasons. First, it plays an essential role in all the above probabilistic control frameworks, either as a modeling choice or derived as an optimal control distribution from a control problem. Second, sampling $\text{exp}(-V(\theta)/\gamma)$ and optimization $\min_\theta V(\theta)$ are dual problems for a general objective function $V(\theta)$ \citep{trillos2023optimization}. When a gradient-based algorithm is considered, sampling can be interpreted as optimization over probability space. From this perspective, sampling control sequences from $p(\mathbf{u}) \propto \text{exp}(-J(\mathbf{u})/\gamma)$ and computing optimal control sequence via $\min_\mathbf{u} J(\mathbf{u})$ are naturally dual problems. Third, the Gibbs distribution is a basis for energy-based generative models \citep{carbone2024hitchhiker}. Therefore, the control-Gibbs distribution draws a connection to generative models for control. Finally, this work is also inspired by the question in behavioral theory \citep{willems1991paradigms}: What is the right model? In this case, we seek the ``right" probabilistic model for a control distribution. This motivates us to place this modeling choice at the central location, and interpret it using different problem formulations that lead to the model.

Besides its conceptual importance, the control-Gibbs distribution also has practical significance: either modeling diverse behaviors or facilitating solving deterministic optimal control problem. We demonstrate the two aspects via projected Langevin dynamics to sample $p(\mathbf{u})$. By generalizing the non-asymptotic analysis of \citep{lamperski2021projected} from static optimization to the control setting, we show that this algorithm can generate diverse controls (i.e., sample the distribution) and solve deterministic optimal control problem (i.e., the solutions converge to a near-optimal control).

Our first contribution is conceptual and review-oriented. The importance of the Gibbs distribution for control is observed in \citep{li2025unifying} to unify update rules underlying MPPI, policy gradient, and diffusion sampling. Complementing this algorithmic view, we take a modeling perspective. We propose to use the control-Gibbs distribution as a fundamental model to connect different probabilistic control frameworks, and we review its key properties. Our second contribution is the non-asymptotic analysis of the projected Langevin sampling for the control-Gibbs distribution.

\section{Control-Gibbs Distribution}
\subsection{Introduction}
Given a nonlinear system $x_{t+1} = f(x_t,u_t)$, with state $x_t \in \mathbb{R}^n$ and input $u_t \in \mathcal{U} \subseteq \mathbb{R}^m$, we consider a standard optimal control cost $J(\mathbf{u};x): \mathcal{U}^{T}\times \mathcal{X} \to \mathbb{R}_{\geqslant 0}$ at any state $x \in \mathcal{X} \subseteq \mathbb{R}^n$:
\begin{align}
J(\mathbf{u};x):= & \sum_{t=0}^{T-1} l(x_t,u_t) + \phi(x_T), \label{eq:OC} \\
& \text{s.t. }  x_{t+1} = f(x_t,u_t), x_0 =x, \nonumber 
\end{align}
with $\mathbf{u} = [u_0^\top,u_{1}^\top,\dots,u_{T-1}^\top]^\top$ and $T$ the horizon. Objective $J$ incorporates the dynamics for notational compactness. For simplicity of exposition, we first let $\mathcal{U} = \mathbb{R}^m$ and $\mathcal{X} =\mathbb{R}^n$. We will explicitly incorporate input constraints when studying a specific sampling algorithm in Section~\ref{sec:Langevin}.

The optimal control problem $\min_{\mathbf{u}} J(\mathbf{u};x)$ models how to select control given the current state: Selecting the best one with respect to an objective. In probabilistic control, we need a probabilistic model to allocate probability mass over all possible controls. We \textit{choose} to focus on the control-Gibbs distribution with the following density $p_\star$, and we will justify this choice in Section~\ref{sec:interpretation}:
\begin{equation} \label{eq:gibbs}
p_\star(\mathbf{u} \mid x) = z^{-1}(x) e^{-J(\mathbf{u};x) /\gamma}, 
\end{equation}
with $\gamma >0$ a temperature parameter and $z(x)$ the normalization constant $z(x) = \int e^{- J(\mathbf{u};x)/\gamma } d\mathbf{u}.$ In this work, all normalizing constants are assumed to exist. Given $x$, a control sequence with a lower cost $J$ is more likely, based on \eqref{eq:gibbs}. When there is a prior or reference density $\bar{p}$, the following Gibbs density is often used:
\begin{equation} \label{eq:genera-gibbs}
p_\textbf{g}(\mathbf{u} \mid x) \propto \bar{p}(\mathbf{u} \mid x) e^{-J(\mathbf{u};x) /\gamma}, 
\end{equation}
where $\bar{p}(\mathbf{u} \mid x)$ adjusts the probability of a given control sequence.

As discussed in \citep{trillos2023optimization}, the optimization problem $ \min_\theta V(\theta)$ and the sampling problem $\theta \sim \pi(\theta) \propto \exp (-V(\theta)/\gamma) $ are dual for a general objective $V(\theta)$. We argue that sampling the control-Gibbs distribution and solving the optimal control problem should be regarded as two analogous problems:
\begin{equation} \label{eq:exp}
\mathbf{u}  \sim p_\star(\mathbf{u} \mid x) \propto  e^{-J(\mathbf{u};x) /\gamma}, \quad  \quad \quad   \min_{\mathbf{u} } J(\mathbf{u};x).
\end{equation}
The optimal control problem seeks a single optimal control sequence. Instead, sampling $p_\star(\mathbf{u} \mid x)$ seeks diverse optimal and near-optimal control sequences, e.g., to capture human behavior variability. The sampling problem can also be regarded as an intermediate step to solve the optimal control problem as $\gamma \to 0$, following simulated annealing. However, our primary interest is in probabilistic control with a non-zero $\gamma$.

\subsection{Interpretation} \label{sec:interpretation}
Why is \eqref{eq:gibbs} a reasonable model for a control distribution? We first note its foundation on optimal control, as it incorporates the control objective $J$. It can also be justified as the \textit{optimal control distribution} of some criterion, as in various probabilistic control frameworks. 

It is the optimal control distribution that minimizes the Helmholtz Free Energy \citep{carbone2024hitchhiker}:  
\begin{equation} \label{eq:free-energy}
  \min_{p } \mathbb{E}_p( J(\mathbf{u};x)) - \gamma \mathcal{H}(p),
\end{equation}
where $\mathcal{H}(p) = - \mathbb{E}_p(\log p(\mathbf{u}\mid x))$ is the entropy. Therefore, it balances decreasing the mean control cost and increasing control uncertainty, e.g., for exploration. If $\gamma = 0$, the solution will degenerate to the deterministic optimal control. Problem~\eqref{eq:free-energy} is a special case of entropy-regularized optimal control \citep{bhole2025unifying} and is also the objective of maximum-entropy RL \citep{haarnoja2018soft}. While the control-Gibbs distribution considers a control sequence, maximum-entropy RL aims to compute a marginal control distribution for closed-loop control based on a variant of dynamic programming, as discussed in Section~\ref{sec:DP}.

The control-Gibbs distribution also arises from the maximum entropy principle \citep{carbone2024hitchhiker}:
\[ 
\max_p \mathcal{H}(p), \text{ s.t.} \int p(\mathbf{u}) d \mathbf{u} =1, \text{ } \mathbb{E}_p(J( \mathbf{u};x)) =J^\star,  
\]
for a constant $J^*$. The optimal unnormalized density is $\text{exp}(-J( \mathbf{u};x)/\gamma(x))$, where the temperature depends on $x$ in general. This dependence can be eliminated by letting $J^\star$ depend on $x$ in a specific way, i.e., $J^\star(x) = \mathbb{E}_{p_\star}(J( \mathbf{u};x))$. We can regard the choice of a constant temperature as a model simplification. Therefore, choosing the control-Gibbs distribution injects minimal information about the distribution, subject to a finite mean cost. This motivation is considered in the maximum entropy control \citep{ziebart2008maximum}.

The Gibbs density with a prior $p_\textbf{g}(\mathbf{u} \mid x) \propto \bar{p}(\mathbf{u} \mid x) e^{-J(\mathbf{u};x) /\gamma}$ is the optimal distribution of the following problem
\begin{equation} \label{eq:MPPI}
\min_{p } \mathbb{E}_p( J(\mathbf{u};x)) + \gamma \mathbb{D}_{\text{KL}}\left(p \,\middle\|\, \bar{p} \right).
\end{equation}
Therefore, it also penalizes the KL divergence between the control distribution and a prior. The control-Gibbs distribution is a special case without a prior. The objective~\eqref{eq:MPPI} is considered in MPPI \citep{williams2018information} and the KL reformulation of path integral control \citep{kappen2016adaptive}. In MPPI, the goal is to compute its mean $\mathbb{E}_{p_\textbf{g}}(\mathbf{u})$ \citep{williams2018information}.

Another probabilistic interpretation is distribution matching \citep{karny1996towards,lefebvre2024probabilistic}, as \eqref{eq:MPPI} can be rewritten to
 \begin{equation} \label{eq:MPPI-matching}
\min_p  \mathbb{D}_{\text{KL}}\left(p \,\middle\|\,   \text{exp}(-\bar{z}^{-1} J(\mathbf{u};x)/\gamma ) \bar{p}  \right),
\end{equation}
with $\bar{z}^{-1}$ a normalization constant so that $  \bar{z}^{-1}\text{exp}(-J(\mathbf{u};x)/\gamma ) \bar{p} $ is a density. In \eqref{eq:MPPI-matching},  we seek a control distribution $p$ to match a target control density. This choice of target distribution appears in \citep{lefebvre2024probabilistic}. It also results as a posterior in the control-as-inference framework after introducing an optimality variable \citep{levine2018reinforcement}. However, the above perspectives leave the choice of target distribution and the optimality variable unjustified. Eq.~\eqref{eq:MPPI-matching} also proves that the control-Gibbs distribution is optimal in \eqref{eq:free-energy} and \eqref{eq:MPPI}, as it leads to zero KL divergence.

 Based on the above, the control-Gibbs distribution plays a central role in different probabilistic control frameworks. We find it conceptually simpler to directly start at \eqref{eq:gibbs} as a modeling choice for probabilistic control. Then sampling the control-Gibbs distribution is a probabilistic counterpart of solving the deterministic optimal control problem. This perspective separates the choice of control distribution from the problem formulations that motivate it. It also aligns with the use of the Gibbs distribution in energy-based models for generative modeling \citep{carbone2024hitchhiker} and for control \citep{urain2025survey}, where inverse problems are often considered, i.e., the energy function is unknown but estimated from data.
 
\section{Closed-Loop Control Distribution: Receding Horizon and Dynamic Programming} \label{sec:DP}
Since we regard the control-Gibbs distribution as the probabilistic analog of the deterministic optimal control problem, it leads to probabilistic extensions of the receding-horizon control and dynamic programming. This allows us to realize closed-loop control.  
 
  A receding-horizon implementation of $p_\star(\mathbf{u}\mid x)$ applies only the first control input after sampling the control sequence. This is equivalent to applying the marginal distribution with density
\begin{equation} \label{eq:softIter}
p_0(u \mid x ) = \int \dots \int p_\star(\mathbf{u} \mid x) du_1 \dots du_{T-1}
\end{equation}
as a closed-loop control distribution. Eq.~\eqref{eq:softIter} can be computed recursively backward from $k=T$ as follows, analogous to dynamic programming.

Define $\mathbf{u}_k = [u_k^\top,\dots,u_{T-1}^\top]^\top$ and the intermediate objective
\begin{align}
J_k(\mathbf{u}_k;x):= & \sum_{t=k}^{T-1} l(x_t,u_t) + \phi(x_T), \text{s.t. }  x_{t+1} = f(x_t,u_t), x_k =x. \nonumber 
\end{align}
Then \eqref{eq:softIter} leads to
\[
p_0(u \mid x ) \propto  e^{-l(x_0,u_0)/\gamma}\dots\int_{u_{T-2}} e^{-l(x_{T-2},u_{T-2})/\gamma} \left( \int_{u_{T-1}} \underbrace{e^{-J_{T-1}(u_{T-1} )/\gamma}}_{p_{T-1} \propto} d u_{T-1} \right)  d u_{T-2}\dots,
\]
which allows a recursive computation from $p_{T-1}$ to $p_0$ as follows, i.e., the so-called soft dynamic programming \citep{haarnoja2017reinforcement,levine2018reinforcement}. 
Define \textit{soft (state-action) value functions} and an operator $\mathcal{F}$ as 
\begin{align}\label{eq:soft-Qdef}
\Tilde{V}_k &:=\mathcal{F}_{\mathbf{u}_k} \{J_k(\mathbf{u}_k;x )\}:=  -\gamma\log \int e^{- J_k(\mathbf{u}_k;x) /\gamma }d \mathbf{u}_k \\
\Tilde{Q}_k(x,u) &:=  l(x,u) + \Tilde{V}_{k+1}(f(x,u)), \nonumber
\end{align}
which satisfies $\Tilde{V}_k = \mathcal{F}_u\{\Tilde{Q}_k(x,u)\}$, where the subscript of $\mathcal{F}$ denotes the integration variable. Therefore, the receding-horizon distribution \eqref{eq:softIter} satisfies
\begin{equation} \label{eq:MPC-distribution}
p_0(u \mid x ) \propto \text{exp}(-\Tilde{Q}_0(x,u)/\gamma) .
\end{equation}
The soft value functions satisfy a variant of the Bellman equation
\begin{align} \label{eq:soft-Bellman}
\Tilde{V}_k(x) &= \mathcal{F}_{u} \left\{  l(x,u)+ \Tilde{V}_{k+1}(f(x,u)) \right\},\\
\Tilde{Q}_k(x,u) &= l(x,u) + \mathcal{F}_{u_+} \{\Tilde{Q}_{k+1}(f(x,u),u_+)\}, \nonumber
\end{align}
with $\Tilde{V}_T(x) =\phi(x)$. The recursion \eqref{eq:soft-Bellman} thus leads to $p_0(u \mid x ) $ via \eqref{eq:MPC-distribution}.

Eq.\eqref{eq:soft-Bellman} extends the classical Bellman equation. The operator $\mathcal{F}$ in \eqref{eq:soft-Qdef} is called \textit{soft minimization} \citep{blondel2024elements}, as an approximation of the $\min$ operator:
\begin{equation} 
   \lim_{\gamma \to 0} -  \gamma \log \int e^{- g(z) / \gamma} dz = \min_z g(z). 
\end{equation}
under some regularity conditions on a function $g(z)$. Therefore, as $\gamma \to 0$, \eqref{eq:soft-Bellman} recovers the standard Bellman equation, and \eqref{eq:soft-Qdef} recovers the standard $Q$ function. 

The soft Bellman equation~\eqref{eq:soft-Bellman} can be equivalently written in a variational form:
\begin{equation} \label{eq:soft-DP_Variational}
\Tilde{V}_k(x_t) = \min_{p(u_t \mid x_t)} \left\{ \mathbb{E}_p \left(l(x_t,u_t)+ \Tilde{V}_{k+1}(x_{t+1}) \right) - \gamma \mathcal{H}(p) \right\}
\end{equation}
due to the Gibbs variational principle \citep[Lemma~4.10]{van2014probability}:
\[\
-  \gamma \log \int e^{- g(z) / \gamma} dz = \min_{p(z)} \mathbb{E}_p(g(z) ) - \gamma \mathcal{H}(p).
\]
The variational form  \eqref{eq:soft-DP_Variational} is used, e.g., in \citep{lambert2026entropy}. The above also shows that the soft value function \eqref{eq:soft-Qdef} represents a minimal entropy-regularized average cost.

\section{Langevin Sampling for Optimal Control and Diverse Actions} \label{sec:Langevin}
Sampling the control-Gibbs distribution can serve dual purposes of both optimization and sampling. To illustrate these two aspects theoretically, instead of advanced sampling approaches such as in \citep{pan2024model}, we consider the fundamental Langevin sampling that allows us to establish non-asymptotic guarantees:
\begin{equation} \label{eq:Langevin}
    \mathbf{u}^{(i+1)} =\Pi_{\mathcal{U}^{T}} \left( \mathbf{u}^{(i)} - \eta \nabla_\mathbf{u}  J(\mathbf{u}^{(i)};x) + \sqrt{2 \eta \gamma}   w^{(i)} \right),
\end{equation}
where $w^{(i)} \sim \mathcal{N}(0, I)$ is i.i.d., $\eta$ is a step size, and initialization $\mathbf{u}^{(0)} \in\mathcal{U}^{T}$ is independent of $w$. The Euclidean projection $\Pi$ ensures the input constraint. The gradient can be computed by the classical adjoint state method in optimal control:
\begin{enumerate}
    \item Forward pass: Obtain state trajectory $\mathbf{x}^{(i)}$ by simulating $x_{k+1} = f(x_k,u_k)$ using $\mathbf{u}^{(i)}$
    \item Backward pass: Obtain costate trajectory $\mathbf{\lambda}^{(i)}$ by simulating \eqref{eq:backward}  backward with the terminal condition $\lambda_T^{(i)} = \nabla_x \phi(x^{(i)}_T )$
    \begin{equation} \label{eq:backward}
\lambda_{k}^{(i)} =  \nabla_x l(x^{(i)}_k,u^{(i)}_k ) +
    \left[\mathbf{J}_x f(x^{(i)}_k,u^{(i)}_k ) \right]^\top  \lambda_{k+1}^{(i)}     \end{equation}

    \item Compute gradient: $\nabla_{u_k}  J(\mathbf{u}^{(i)};x) =\nabla_u l(x^{(i)}_k,u^{(i)}_k ) + \left[\mathbf{J}_u f(x^{(i)}_k,u^{(i)}_k ) \right]^\top \lambda_{k+1}^{(i)}$
\end{enumerate}
The above algorithm is a stochastic variant of the classical gradient-based approach for optimal control \cite[Chap. 7.4]{bryson1975applied}, with the additional noise $w $.

Sampling for optimization concerns solving the deterministic optimal control problem, i.e., finding a control input sequence that is optimal or near-optimal. For static optimization, sampling helps escape local optima and obtain a near-optimum solution, even if the objective is non-convex \citep{raginsky2017non}. This also holds with a constant non-zero temperature, without requiring $\gamma \to 0$.

Sampling itself leads to diverse control within the control distribution. Theoretically, it concerns whether the sampling algorithm can cover the whole distribution, while practically, optimal or near-optimal solutions are more likely to be generated due to finite iterations of the Langevin dynamics.

The dual aspects of \eqref{eq:Langevin} for optimization and sampling can be established theoretically. We extend the non-asymptotic analysis in \citep{lamperski2021projected} from static optimization to the control setting. The difference from \citep{lamperski2021projected} is that the gradient in \eqref{eq:Langevin} has a control structure, which requires suitable regularity conditions.

\begin{ass} \label{ass:BoundRegion}
  The input constraint $\mathcal{U}$ is compact, convex and contains an open ball around the origin. The initial state region $\mathcal{X}$ is compact.
\end{ass}

\begin{ass}\label{ass:Lips} The functions $f$, $l$, and $\phi$ are in $\mathcal{C}^1$. Their gradients $\nabla_{(x,u)}f$, $\nabla_{(x,u)} l $, and $ \nabla_{x} \phi$ are locally Lipschitz continuous in $(x,u)$.
\end{ass}

Then the sampling perspective of \eqref{eq:Langevin} concerns covering diverse solutions, i.e., whether the distribution of $\mathbf{u}^{(N)}$, denoted by $P_N$, converges to the control-Gibbs distribution $P_\star$ with density $p_\star \propto \mathbf{1}_{\mathcal{U}^T}(\mathbf{u}) \text{exp}(-J(\mathbf{u};x)/\gamma) $, where $\mathbf{1}$ denotes the indicator function. Proposition~\ref{Prop:Optimization} provides this convergence guarantee.

\begin{prop}[Sampling] \label{Prop:Optimization}
Let Assumptions~\ref{ass:BoundRegion} and \ref{ass:Lips} hold, and consider any $x \in \mathcal{X}$. There exist positive constants $a$, $c_1$, and $c_2$ such that if $\eta = \log N/ (4a N) \leqslant 1/2$ and $N \geqslant 4$, then
$$
W_1(P_N, P_\star) \leqslant \left(
c_1 + \frac{c_2}{(4a)^{1/4}}\right) N^{-1/4}\left( \log N \right)^{1/2}. 
$$
where $W_1$ denotes the Wasserstein $1$-distance.
\end{prop} 

Finally, Proposition~\ref{Prop:Sampling} demonstrates the optimal control perspective. It shows sampling leads to near-optimal control sequences, up to an $\varepsilon$ error in the control objective.  

 \begin{prop}[Optimal Control] \label{Prop:Sampling}
Let Assumptions~\ref{ass:BoundRegion} and \ref{ass:Lips} hold, and consider any $x \in \mathcal{X}$. There exist positive constants $a$, $c_3$, $c_4$, and $c_5$ such that for any $\varepsilon>0$, it holds 
$$
\mathbb{E}(J(\mathbf{u}^{(N)};x) ) \leqslant \min_{\mathbf{u} \in \mathcal{U}^{T}} J(\mathbf{u};x) +\varepsilon,
$$
if $\eta = \log N /(4a N )  \leqslant 1/2$, $N \geqslant \max\{4, \frac{1}{\varepsilon^5} \left( 10c_4 \right)^{5/2} e^{c_5 /\gamma } \}$, and $\gamma \leqslant c_3 \left( \frac{  \varepsilon  c_3 e}{4 Tm } \right)^2$.
\end{prop} 

\section{Numerical Example: Arm Trajectory Generation}
An arm trajectory in $2$D space can be modeled by a trajectory of a $2$D point that minimizes jerk over time \citep{flash1985coordination}. We use this example to illustrate the control-Gibbs distribution for diverse trajectory generation and for optimization. Consider sampling trajectories from position $(0,0)$ to $(1,0)$ while avoiding an obstacle point and minimizing jerk. The continuous-time dynamics is $\dot{q}_t = v_t$, $\dot{v}_t = a_t$, and $\dot{a}_t = u_t$, where input $u_t \in \mathbb{R}^2$ is jerk, and $q$, $v$, $a$ denote the position, velocity, and acceleration, respectively. To be consistent with the rest of this work, we consider its discrete-time model based on the zero-order hold, with sampling interval $h = 1/6$.

 The control objective is 
\[
J(\mathbf{u})
=
  \sum_{k=0}^{T-1}
\left[
\|u_k\|_2^2
+
2 \text{exp}\!\left(
-\frac{\|q_{k+1}-c\|_2^2}{2\sigma^2}
\right)
\right],
\] 
where $T=24$ covers $4$ seconds, $c \in \mathbb{R}^2$ is the obstacle position, and $\sigma=0.1$. The terminal constraint $q_T = (1,0)^\top$ is eliminated as follows. We consider the initial state at the origin, then $q_T = \mathcal{C} \mathbf{u}$ with $ \mathcal {C} $ containing two rows of the controllability matrix. We re-parameterize $\mathbf{u} =\bar{\mathbf{u}}  + Q \mathbf{z}$, where $\bar{\mathbf{u}} = \mathcal{C}^\top(\mathcal{C}\mathcal{C}^\top )^{-1}q_T$ is a nominal solution of $q_T = \mathcal{C} \mathbf{u}$ and $Q$ satisfies $\mathcal{C} Q = 0$ and $Q^\top Q = I$. Consider $\bar{J}(\mathbf{z}):= J(\bar{\mathbf{u}}  + Q \mathbf{z})$, and we sample $p_\star \propto \text{exp}(-\bar{J}(\mathbf{z})/\gamma)$ using \eqref{eq:Langevin} with $\eta=0.01$. For simplicity, the input bound is set to be large enough so that it never activates.

\begin{figure}[h]
    \centering
    \includegraphics[width=\linewidth]{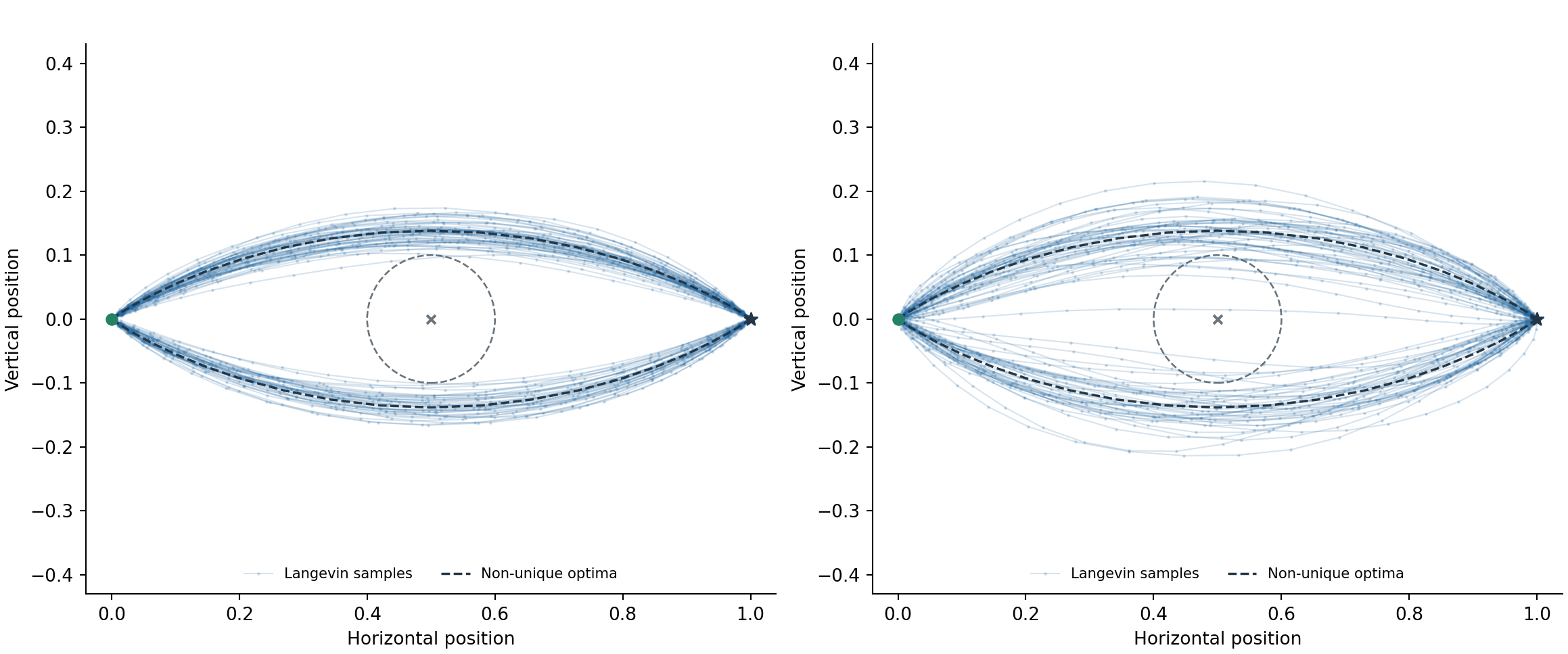}
    \caption{Arm trajectories with temperature $\gamma=0.12$ (left) and $\gamma=0.48$ (right) after $i \geqslant 2000$. The cross is the obstacle position $(0.5,0)$, and the radius of the dashed circle is $\sigma$.}
    \label{fig:1}
\end{figure}

To illustrate diverse solutions, we first choose $c = (0.5,0)^\top$ to create non-unique optimal input sequences that are numerically found, leading to the two dashed trajectories in Figure~\ref{fig:1}. Diverse trajectories can be sampled around the optima and are affected by temperature $\gamma$. 

\begin{figure}[h]
    \centering
    \includegraphics[width=\linewidth]{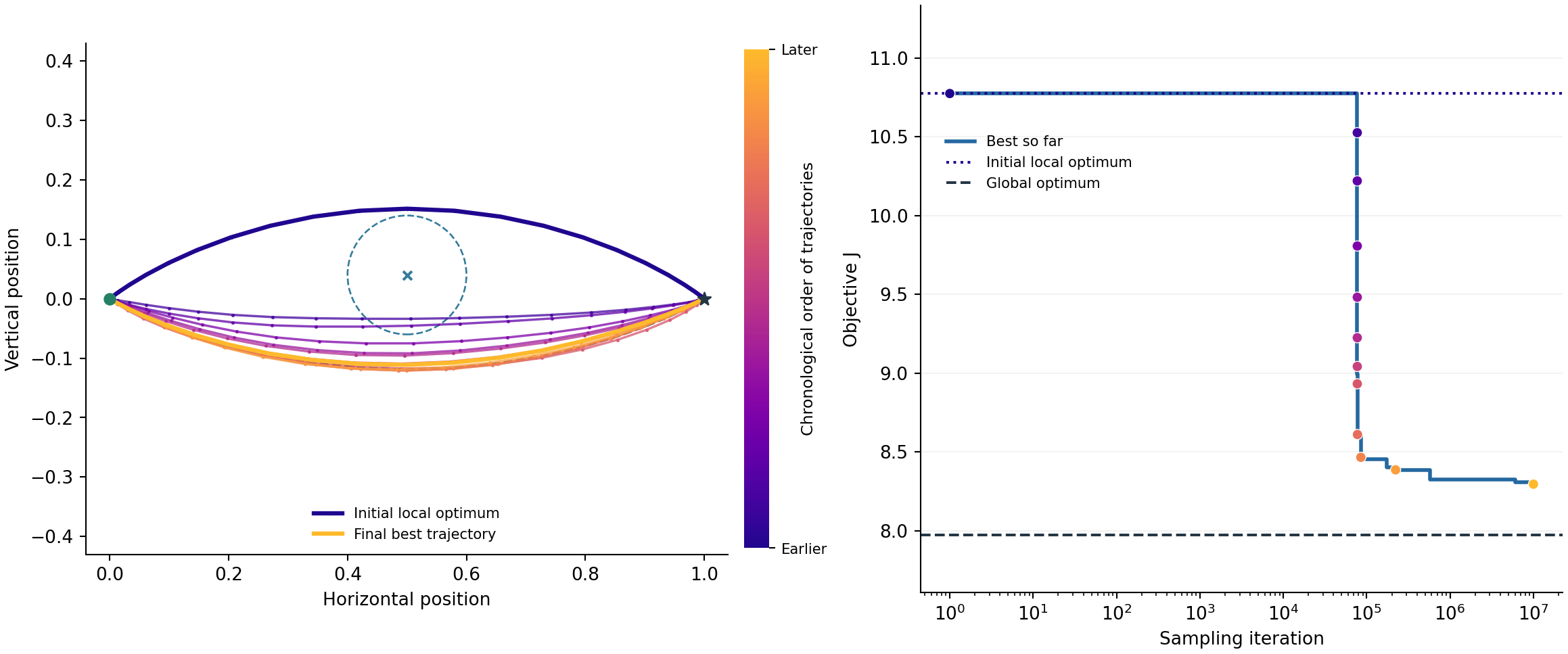}
    \caption{Trajectories (best so far) under $\gamma =0.06$ can escape the initialized local optimum (right) with a suboptimality gap.}
    \label{fig:2}
\end{figure}

For optimization, we let the obstacle position $c = (0.5,0.04)$, so that the optimal trajectory is unique, i.e., a trajectory below the obstacle. However, a local optimum above the obstacle might trap the algorithm. As shown in Figure~\ref{fig:2}, we initialize \eqref{eq:Langevin} at the local optimum, and the sampling algorithm escapes it and converges closer to the optimum, with a suboptimality gap. Note that at the shown iterations, the best trajectories among all the previous samples are ploted in Figure~\ref{fig:2}.
\section{Proof of Propositions~\ref{Prop:Optimization} and \ref{Prop:Sampling}}
The two propositions extend Theorems~1 and Proposition~3 from \citep{lamperski2021projected} from a general energy function to a control structure. The results in \citep{lamperski2021projected} consider the projected Langevin sampling
  $$ \theta^{(i+1)} =\Pi_{\Theta} \left( \theta^{(i)} - \eta \nabla_\theta  V(\theta^{(i)},z_i) + \sqrt{2 \eta \gamma}   w^{(i)} \right),
$$
for an unnormalized Gibbs density $\text{exp}(- \mathbb{E}_z(V(\theta,z) )/\gamma)$, where $z$ is random, and $z_i$ is i.i.d over $k$ with identifical distribution as $z$. In addition, $\nabla_\theta  V(\theta,z) - \nabla_\theta  \mathbb{E}_z(V(\theta,z) )$ should be sub-Gaussian for any $\theta$. Note that for any fixed $x \in \mathcal{X}$, \eqref{eq:Langevin} satisfies the above setting, when $z$ and $z_i$ are replaced by $x$ with the identical and independent Dirac delta distribution $\delta_{x}$. In addition, $\nabla_\mathbf{u}  J(\mathbf{u};x) - \nabla_\mathbf{u}  \mathbb{E}_{x} J(\mathbf{u};x)$ is constant zero and is thus sub-Gaussian.

Finally, the results in \citep{lamperski2021projected} hold if the gradient $\nabla_\mathbf{u}  J(\mathbf{u};x) $ is Lipschitz continuous in $\mathbf{u} \in \mathcal{U}^{T}$ for every $x \in \mathcal{X}$ and is also Lipschitz in $x \in \mathcal{X}$ for every $\mathbf{u} \in \mathcal{U}^{T}$. These properties hold if $\nabla_\mathbf{u}  J(\mathbf{u};x)$ is jointly Lipschitz on $\mathcal{U}^{T} \times \mathcal{X}$, which can be established under the considered regularity conditions on the control problem, as shown in Lemma~\ref{lem:Lipschitz}. Note that Prop.~\ref{Prop:Sampling} results from \citep[Prop. 3]{lamperski2021projected} with its parameters chosen as $\rho=5$ and $\zeta = 2$ here.
\begin{lem} \label{lem:Lipschitz}
Given any initial states $x_0^{(1)}$,$x_0^{(2)}  \in \mathcal{X}$ and corresponding input sequences $\mathbf{u}^{(1)}$, $\mathbf{u}^{(2)} \in \mathcal{U}^{T}$, consider the resulting state sequences $\mathbf{x}^{(1)}$, $\mathbf{x}^{(2)}$ and the associated co-state trajectories $\mathbf{\lambda}^{(1)}$, $\mathbf{\lambda}^{(2)}$. If Assumptions~\ref{ass:BoundRegion} and \ref{ass:Lips} hold, there exist positive constants $L_x$, $P_x$, $L_\lambda$, and $P_\lambda$ such that for all $k=0,\dots,T$,
\begin{align} 
    \|x_k^{(1)} - x_k^{(2)} \|& \leqslant L_x  \|\mathbf{u}^{(1)} - \mathbf{u}^{(2)} \|+ P_x  \|x_0^{(1)} - x_0^{(2)} \|, \label{eq:StateDeviation}\\
     \|\lambda_k^{(1)} - \lambda_k^{(2)}\| & \leqslant L_\lambda \|\mathbf{u}^{(1)} - \mathbf{u}^{(2)} \| + P_\lambda  \|x_0^{(1)} - x_0^{(2)} \|. \label{eq:CoStateDeviation}
\end{align}
Moreover, $\nabla_\mathbf{u}  J(\mathbf{u};x)$ is Lipschitz on $\mathcal{U}^{T} \times \mathcal{X}$.
\end{lem}
\begin{proof}
Let $\lesssim$ denote inequality up to a constant. Given that $f$ is in $\mathcal{C}^1$ and Assumption~\ref{ass:BoundRegion} holds, any resulting state sequence of a given horizon $T$ will be in some compact region, denoted by $\mathcal{R}$. Assumption~\ref{ass:Lips} implies that $f$, $l$, $\phi$, and their gradients are Lipschitz within the compact region $\mathcal{U} \times \mathcal{R}$. Then, recursively applying the Lipschitz property of $f$ on $\mathcal{U} \times \mathcal{R}$ proves \eqref{eq:StateDeviation}. For \eqref{eq:CoStateDeviation}, as any state and input sequences belong to the compact region $\mathcal{U} \times \mathcal{R}$, Assumption~\ref{ass:Lips} and the backward pass \eqref{eq:backward} imply that $\mathbf{\lambda}$ and $\mathbf{J}f(x _k,u_k) $ are uniformly bounded, i.e., $\| \lambda_k \| \leqslant c_{\text{max}}$ and $\|\mathbf{J}f(x _k,u_k)\| \leqslant j_{\text{max}}$ for any $k \leqslant T$. Then, we prove \eqref{eq:CoStateDeviation} by induction, backward starting from $k=T$. Given \eqref{eq:StateDeviation} and that $\nabla \phi$ is Lipschitz continuous on $\mathcal{R}$ , $\|\lambda_T^{(1)} - \lambda_T^{(2)}\| = \|\nabla \phi(x_T^{(1)})- \nabla \phi(x_T^{(2)}) \| \lesssim \|\mathbf{u}^{(1)} - \mathbf{u}^{(2)} \|+ \|x_0^{(1)} - x_0^{(2)} \|$. Assume $\|\lambda_{k+1}^{(1)} - \lambda_{k+1}^{(2)}\|\lesssim \|\mathbf{u}^{(1)} - \mathbf{u}^{(2)} \|+ \|x_0^{(1)} - x_0^{(2)} \|$, then $\|\lambda_k^{(1)} - \lambda_k^{(2)}\|\lesssim  \| \nabla_x l(x^{(1)}_k,u^{(1)}_k ) - \nabla_x l(x^{(2)}_k,u^{(2)}_k ) \| + \|\mathbf{J}_x f(x^{(1)}_k,u^{(1)}_k ) \|\|\lambda^{(1)}_{k+1}-\lambda^{(2)}_{k+1}\| + \| \lambda^{(2)}_{k+1}\| \|\mathbf{J}_x f(x^{(1)}_k,u^{(1)}_k )-\mathbf{J}_x f(x^{(2)}_k,u^{(2)}_k ) \| \lesssim \| \nabla_x l(x^{(1)}_k,u^{(1)}_k ) - \nabla_x l(x^{(2)}_k,u^{(2)}_k ) \| + j_{\text{max}}\|\lambda^{(1)}_{k+1}-\lambda^{(2)}_{k+1}\| + c_{\text{max}} \|\mathbf{J}_x f(x^{(1)}_k,u^{(1)}_k )-\mathbf{J}_x f(x^{(2)}_k,u^{(2)}_k ) \| \lesssim \|[x^{(1)}_k;u^{(1)}_k] - [x^{(2)}_k;u^{(2)}_k] \|+ \|\lambda^{(1)}_{k+1}-\lambda^{(2)}_{k+1}\| \lesssim \|\mathbf{u}^{(1)} - \mathbf{u}^{(2)} \|+ \|x_0^{(1)} - x_0^{(2)} \|$, which concludes the induction.

Finally, to prove the Lipschitz continuity of $\nabla_\mathbf{u}  J(\mathbf{u};x)$ on $\mathcal{U}^{T} \times \mathcal{X}$, we have the following based on the adjoint state method:
 \begin{align*}
 & \| \nabla_{u_k} J(\mathbf{u}^{(1)};x_0^{(1)}) -  \nabla_{u_k} J(\mathbf{u}^{(2)};x_0^{(2)})\| \\
 \leqslant & \|\nabla_u l(x^{(1)}_k,u^{(1)}_k ) -  \nabla_u l(x^{(2)}_k,u^{(2)}_k ) \| \\
 & + \|\left[\mathbf{J}_u f(x^{(1)}_k,u^{(1)}_k ) \right]^\top \lambda_{k+1}^{(1)}-\left[\mathbf{J}_u f(x^{(2)}_k,u^{(2)}_k ) \right]^\top \lambda_{k+1}^{(2)} \|   \\
 \stackrel{(a)}{\lesssim} &   \sqrt{\|x^{(1)}_k - x^{(2)}_k\|^2 +\|u^{(1)}_k - u^{(2)}_k\|^2 } \\
 & +  \| \mathbf{J}_u f(x^{(1)}_k,u^{(1)}_k )   -  \mathbf{J}_u f(x^{(2)}_k,u^{(2)}_k )    \| \|\lambda_{k+1}^{(1)}\| \\
 & + \|\mathbf{J}_u f(x^{(2)}_k,u^{(2)}_k )\| \|  \lambda_{k+1}^{(1)}  -  \lambda_{k+1}^{(2)} \|  \\
  \stackrel{(b)}{\lesssim} & \sqrt{\|x^{(1)}_k - x^{(2)}_k\|^2 +\|u^{(1)}_k - u^{(2)}_k\|^2 }+      \|  \lambda_{k+1}^{(1)}  -  \lambda_{k+1}^{(2)} \|      \\
 \stackrel{(c)}{\lesssim} &    \|\mathbf{u}^{(1)} - \mathbf{u}^{(2)} \| +  \|x_0^{(1)} - x_0^{(2)} \|,
 \end{align*}
 where step $(a)$ uses the Lipschitz continuity of $\nabla_u l$ on the compact region $\mathcal{U} \times \mathcal{R}$; step $(b)$ follows from the Lipschitz continuity of $\mathbf{J} f$ on the compact region and \eqref{eq:StateDeviation}; and step $(c)$ uses \eqref{eq:StateDeviation} and \eqref{eq:CoStateDeviation}. Therefore, we have $   \|\mathbf{u}^{(1)} - \mathbf{u}^{(2)} \| +  \|x_0^{(1)} - x_0^{(2)} \| \gtrsim \sum_{k=0}^{T-1} \| \nabla_{u_k} J(\mathbf{u}^{(1)};x_0^{(1)}) -  \nabla_{u_k} J(\mathbf{u}^{(2)};x_0^{(2)})\| \geqslant  \| \nabla_{\mathbf{u}} J(\mathbf{u}^{(1)};x_0^{(1)}) -  \nabla_{\mathbf{u}} J(\mathbf{u}^{(2)};x_0^{(2)})\| $, proving the result.
\end{proof}
\section{Conclusion}
The control-Gibbs distribution is a fundamental model in various probabilistic control frameworks. It is an optimal control distribution that maximizes entropy and minimizes regularized average control cost. A probabilistic variant of dynamic programming also allows computing marginal distributions for closed-loop control. We argue that sampling the control-Gibbs distribution should be regarded as an analogous problem to solving a deterministic optimal control problem. In this work, we provide non-asymptotic guarantees for the projected Langevin sampling algorithm. These guarantees show that sampling the control-Gibbs distribution enables diverse control solutions and finds near-optimal solution for the deterministic optimal control problem.

\subsection*{Generative AI Use}
 Generative AI tools assisted with coding the numerical example and polishing grammar after the manuscript was finished. The author
reviewed and verified all AI-assisted content and take responsibility for the
final content of the paper.

\bibliography{l4dc2027-sample}

@article{urain2025survey,
  title={A Survey on Deep Generative Models for Robot Learning From Multimodal Demonstrations},
  author={Urain, J. and Mandlekar, A. and Du, Y. and Muhammad, N. and Xu, D. and Fragkiadaki, K. and Chalvatzaki, G. and Peters, J.},
  journal={IEEE Transactions on Robotics},
  volume={42},
  pages={60--79},
  year={2025},
  publisher={IEEE}
}

@article{dhawale2017role,
  title={The Role of Variability in Motor Learning},
  author={Dhawale, A. K. and Smith, M. A. and {\"O}lveczky, B. P.},
  journal={Annual Review of Neuroscience},
  volume={40},
  pages={479--498},
  year={2017},
  publisher={Annual Reviews}
}

@article{chi2025diffusion,
  title={Diffusion policy: Visuomotor policy learning via action diffusion},
  author={Chi, C. and Xu, Z. and Feng, S. and Cousineau, E. and Du, Y. and Burchfiel, B. and Tedrake, R. and Song, S.},
  journal={The International Journal of Robotics Research},
  volume={44},
  number={10-11},
  pages={1684--1704},
  year={2025}
}

@inproceedings{kappen2007introduction,
  title={An introduction to stochastic control theory, path integrals and reinforcement learning},
  author={Kappen, H. J.},
  booktitle={AIP conference proceedings},
  volume={887},
  number={1},
  pages={149--181},
  year={2007}
}

@article{williams2018information,
  title={Information-theoretic model predictive control: Theory and applications to autonomous driving},
  author={Williams, G. and Drews, P. and Goldfain, B. and Rehg, J. M. and Theodorou, E. A.},
  journal={IEEE Transactions on Robotics},
  volume={34},
  number={6},
  pages={1603--1622},
  year={2018},
  publisher={IEEE}
}

@inproceedings{raginsky2017non,
  title={Non-convex learning via stochastic gradient {L}angevin dynamics: a nonasymptotic analysis},
  author={Raginsky, M. and Rakhlin, A. and Telgarsky, M.},
  booktitle={Conference on Learning Theory},
  pages={1674--1703},
  year={2017},
  organization={PMLR}
}

@book{bryson1975applied,
  title={Applied {O}ptimal {C}ontrol: {O}ptimization, {E}stimation, and {C}ontrol},
  author={Bryson, A. E. and Ho, Y. C. },
  publisher={Hemisphere Pub. Corp},
  year={1975}
}

@article{trillos2023optimization,
  title={From optimization to sampling through gradient flows},
  author={Garcia Trillos, N. and Hosseini, B. and Sanz-Alonso, D.},
  journal={Notices of the American Mathematical Society},
  volume={70},
  number={6},
  pages={8},
  year={2023}
}

@article{levine2018reinforcement,
  title={Reinforcement learning and control as probabilistic inference: Tutorial and review},
  author={Levine, S.},
  journal={arXiv preprint arXiv:1805.00909},
  year={2018}
}

@inproceedings{haarnoja2017reinforcement,
  title={Reinforcement learning with deep energy-based policies},
  author={Haarnoja, T. and Tang, H. and Abbeel, P. and Levine, S.},
  booktitle={International Conference on Machine Learning},
  pages={1352--1361},
  year={2017},
  organization={PMLR}
}

@article{bhole2025unifying,
  title={Unifying entropy regularization in optimal control: From and back to classical objectives via iterated soft policies and path integral solutions},
  author={Bhole, A. and Filabadi, M. M. and Crevecoeur, G. and Lefebvre, T.},
  journal={arXiv preprint arXiv:2512.06109},
  year={2025}
}

@article{lefebvre2024probabilistic,
  title={Probabilistic control and majorisation of optimal control},
  author={Lefebvre, T.},
  journal={Systems \& Control Letters},
  volume={190},
  pages={105837},
  year={2024}
}

@article{carbone2024hitchhiker,
  title={Hitchhiker's guide on the relation of Energy-Based Models with other generative models, sampling and statistical physics: a comprehensive review},
  author={Carbone, D.},
  journal={Transactions on Machine Learning Research},
  year={2025}
}

@inproceedings{lamperski2021projected,
  title={Projected stochastic gradient Langevin algorithms for constrained sampling and non-convex learning},
  author={Lamperski, A.},
  booktitle={Conference on Learning Theory},
  pages={2891--2937},
  year={2021},
  organization={PMLR}
}

@article{flash1985coordination,
  title={The coordination of arm movements: an experimentally confirmed mathematical model},
  author={Flash, T. and Hogan, N.},
  journal={The Journal of Neuroscience},
  volume={5},
  number={7},
  pages={1688--1703},
  year={1985}
}

@article{willems1991paradigms,
  title={Paradigms and puzzles in the theory of dynamical systems},
  author={Willems, J. C.},
  journal={IEEE Transactions on Automatic Control},
  volume={36},
  number={3},
  pages={259--294},
  year={1991}
}

@article{karny1996towards,
  title={Towards fully probabilistic control design},
  author={K{\'a}rn{\`y}, M.},
  journal={Automatica},
  volume={32},
  number={12},
  pages={1719--1722},
  year={1996}
  }

@article{lambert2026entropy,
  title={Entropy Regularized Variational Dynamic Programming for Stochastic Optimal Control},
  author={Lambert, M. and Bach, F. and Bonnabel, S.},
  journal={IEEE Transactions on Automatic Control},
  year={2026}
}

@article{kappen2016adaptive,
  title={Adaptive importance sampling for control and inference},
  author={Kappen, H. J. and Ruiz, H. C.},
  journal={Journal of Statistical Physics},
  volume={162},
  number={5},
  pages={1244--1266},
  year={2016},
  publisher={Springer}
}

@article{li2025unifying,
  title={Unifying model predictive path integral control, reinforcement learning, and diffusion models for optimal control and planning},
  author={Li, Y. and Chen, M.},
  journal={arXiv preprint arXiv:2502.20476},
  year={2025}
}

@article{forssell1999closed,
  title={Closed-loop identification revisited},
  author={Forssell, U. and Ljung, L.},
  journal={Automatica},
  volume={35},
  number={7},
  pages={1215--1241},
  year={1999}
}

@inproceedings{athrey2024regret,
  title={Regret analysis of learning-based linear quadratic gaussian control with additive exploration},
  author={Athrey, A. and Mazhar, O. and Guo, M. and De Schutter, B. and Shi, S.},
  booktitle={2024 European Control Conference (ECC)},
  pages={1795--1801},
  year={2024}
}

@inproceedings{haarnoja2018soft,
  title={Soft actor-critic: Off-policy maximum entropy deep reinforcement learning with a stochastic actor},
  author={Haarnoja, T. and Zhou, A. and Abbeel, P. and Levine, S.},
  booktitle={International Conference on Machine Learning},
  pages={1861--1870},
  year={2018}
}

@inproceedings{ziebart2008maximum,
  title={Maximum entropy inverse reinforcement learning},
  author={Ziebart, B. D. and Maas, A. L. and Bagnell, J. A. and Dey, A. K.},
  booktitle={AAAI},
  volume={8},
  pages={1433--1438},
  year={2008}
}

@article{blondel2024elements,
  title={The {E}lements of {D}ifferentiable {P}rogramming},
  author={Blondel, M. and Roulet, V.},
  journal={arXiv preprint arXiv:2403.14606},
  year={2024}
}

@article{pan2024model,
  title={Model-based diffusion for trajectory optimization},
  author={Pan, C. and Yi, Z. and Shi, G. and Qu, G.},
  journal={Advances in Neural Information Processing Systems},
  volume={37},
  pages={57914--57943},
  year={2024}
}

@article{van2014probability,
  title={Probability in {H}igh {D}imension},
  author={Van Handel, R.},
  journal={Lecture Notes (Princeton University)},
  volume={2},
  number={3},
  pages={2--3},
  year={2014}
}

\end{document}